\documentclass[10pt,conference]{IEEEtran}
\IEEEoverridecommandlockouts
\usepackage{enumitem}
\usepackage{cite}
\usepackage{amsmath,amssymb,amsfonts}
\usepackage{algorithmic}
\usepackage{graphicx}
\usepackage{textcomp}
\usepackage{xcolor}
\usepackage{url} 

\def\BibTeX{{\rm B\kern-.05em{\sc i\kern-.025em b}\kern-.08em
    T\kern-.1667em\lower.7ex\hbox{E}\kern-.125emX}}

\newcommand{\linguistic}[1]{{\textit{baseline+psycholinguistic}}}
\newcommand{\morality}[1]{{\textit{baseline+psycholinguistic+morality}}}

\begin{document}

\title{Analyzing Toxicity in Open Source Software Communications Using Psycholinguistics and \\ Moral Foundations Theory}

\author{
\IEEEauthorblockN{Ramtin Ehsani}
\IEEEauthorblockA{Drexel University \\
Philadelphia, PA, USA \\
ramtin.ehsani@drexel.edu}
\and
\IEEEauthorblockN{Rezvaneh (Shadi) Rezapour}
\IEEEauthorblockA{Drexel University \\
Philadelphia, PA, USA \\
shadi.rezapour@drexel.edu}
\and
\IEEEauthorblockN{Preetha Chatterjee}
\IEEEauthorblockA{Drexel University \\
Philadelphia, PA, USA \\
preetha.chatterjee@drexel.edu}
}

\maketitle

\begin{abstract}

Studies have shown that toxic behavior can cause contributors to leave, and hinder newcomers' (especially from underrepresented communities) participation in Open Source Software (OSS) projects. Thus, detection of toxic language plays a crucial role in OSS collaboration and inclusivity. 
Off-the-shelf toxicity detectors are ineffective when applied to OSS communications, due to the distinct nature of toxicity observed in these channels (e.g., entitlement and arrogance are more frequently observed on GitHub than on Reddit or Twitter). 
In this paper, we investigate a machine learning-based approach for the automatic detection of toxic communications in OSS. We leverage psycholinguistic lexicons, and Moral Foundations Theory to analyze toxicity in two types of OSS communication channels; issue comments and code reviews. Our evaluation indicates that our approach can achieve a significant performance improvement (up to 7\% increase in F1 score) over the existing domain-specific toxicity detector. We found that using moral values as features is more effective than linguistic cues, resulting in 67.50\% F1-measure in identifying toxic instances in code review data and 64.83\% in issue comments. While the detection accuracy is far from accurate, this improvement demonstrates the potential of integrating moral and psycholinguistic features in toxicity detection models. These findings highlight the importance of context-specific models that consider the unique communication styles within OSS, where interpersonal and value-driven language dynamics differ markedly from general social media platforms. Future work could focus on refining these models to further enhance detection accuracy, possibly by incorporating community-specific norms and conversational context to better capture the nuanced expressions of toxicity in OSS environments.
\end{abstract}

\begin{IEEEkeywords}
moral principles, toxicity, open source, textual analysis
\end{IEEEkeywords}

\section{Introduction}
\begin{figure}[t] 
\includegraphics[width=1\linewidth]{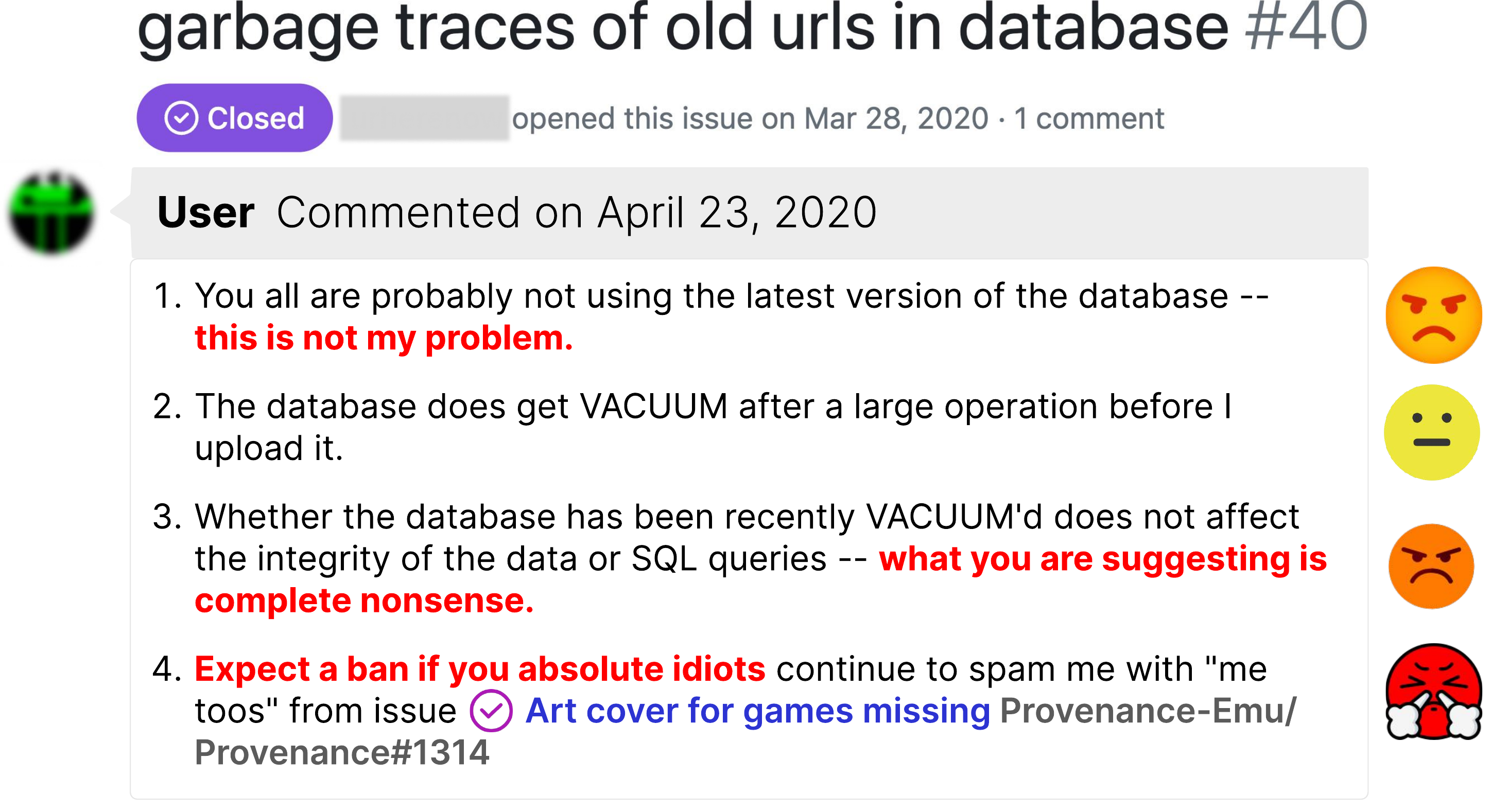}
\vspace{-0.7cm}
\caption{A Toxic Conversation in OSS.}
\label{OSSToxicExample}
\vspace{-0.7cm}
\end{figure}

Open Source Software (OSS) projects are a societal good and they have increased the speed of digital advancement. 
However, most new OSS projects fail and many longstanding projects are abandoned by developers
~\cite{coelho2017why}.
A common reason contributors abandon projects is because of social and emotional factors~\cite{9402044}, e.g., a negative experience with other participants through a toxic conversation (as shown in Figure \ref{OSSToxicExample}). Uncivil language can also deter newcomers from contributing to OSS~\cite{Igor15, qiu19, 10555803}. While some OSS projects implement codes of conduct to define acceptable behavior, manually monitoring adherence is challenging for maintainers due to the high volume of daily communications~\cite{tourani2018code}. 
If project maintainers and participants were able to proactively detect and prevent toxic communications through automated tools, it would lead to more inclusive and sustainable OSS.

There are several challenges in automatically detecting toxic content in OSS. Directly applying toxicity detection tools trained on other domains (e.g., Google Perspective API) to software engineering (SE) corpora has proven ineffective due to the unique language and norms in SE~\cite{Raman2020, Sarker2020ABS}. 
Toxicity in OSS is often nuanced, including insults resulting from technical disagreements and comments that are entitled, demanding, or arrogant~\cite{Miller2022}. Automated toxicity detectors often fail to capture this ``covert toxicity", expressed through cynicism and entitlement in SE communications~\cite{Lees2021CapturingCT, 9793879}.
Recent advances in Large Language Models (LLMs) have spurred interest in their application for toxicity detection on platforms like GitHub~\cite{10.1145/3639476.3639777, 10.1145/3643829}. 
However, LLMs also struggle with SE-specific toxicity (e.g., F-measure of 0.62~\cite{10.1145/3639476.3639777}), particularly when it lacks clear indicators such as offensive words or hate speech. For instance, statements like “\textit{Such a plugin already exists...You have nobody to blame but yourself}” are misclassified, as LLMs may fail to recognize the subtle condescension and irony present in OSS communications.


Beyond Software Engineering, researchers in the Natural Language Processing (NLP) field have developed methods to automatically detect cyberbullying~\cite{cyberbullying}, offensive language~\cite{offensive}, and hate speech~\cite{Vigna2017HateMH, Chaudhary2021CounteringOH}. For example, using (psycho)linguistic markers derived from the Linguistic Inquiry and Word Count (LIWC) dictionary~\cite{LIWC} has proven effective for analyzing toxic language across various communities~\cite{an2021predicting,silva2020data,salminen2020developing}. Studies consistently demonstrate that linguistic styles and word choices offer valuable insights into individuals' thoughts, opinions, and emotions~\cite{triandis1989self,boyd2022development,rezapour2019}.

Moral Foundations Theory (MFT) provides a complementary perspective by proposing that judgments are guided by intuitive appraisals based on personal values~\cite{graham2013moral,haidt2004intuitive}. 
People's values and personal norms affect their (spontaneous) attitude, decision-making process, and what they perceive as good or bad, and moral or immoral \cite{haidt2001emotional,rezapour2021incorporating}. 
MFT categorizes people's moral reactions and behavior into five foundations or principles which are each further characterized by two opposing values as virtues (good) and vices (bad); care/harm, fairness/cheating, authority/subversion, loyalty/betrayal, and purity/degradation \cite{graham2013moral,haidt2004intuitive}. 
Previous research in psychology has addressed the connection between moral foundations and hate and showed that morality is a key feature of hatred; hate is connected to core moral beliefs and higher levels of moral emotions (e.g., contempt, anger, and disgust)~\cite{morality_hate_2018,morality_hate_2021}. 
Ehsani et al. examined the role of morality in OSS toxicity and highlighted connections between moral principles and toxic interactions on GitHub issue threads~\cite{10.1145/3611643.3613077}. Their qualitative study revealed that \textit{purity/degradation} and \textit{care/harm} were the most frequently observed moral principles, often linked to toxic threads that contain insults. 

Building on Ehsani et al's work, we investigate if integrating moral dimensions with psycholinguistic markers can enhance automated toxicity detection in OSS communication. 
Specifically, we explore three machine learning-based classifiers (Support Vector Machine, Logictic Regression, Gradient Boosting) that leverage psycholinguistic lexicons (LIWC~\cite{boyd2022development}) and MFT dimensions~\cite{graham2013moral,haidt2004intuitive} as features to detect toxicity in OSS communications. To ensure generalizability across different communication channels, we detect and quantify expressions of moral values as well as psychological markers in two types of software-related artifacts: (a) GitHub issue comments~(dataset by \cite{Raman2020}), and (b) Gerrit code review comments (extended dataset by \cite{Sarker2020ABS}), to investigate the following research questions:

\begin{itemize}[leftmargin=*]
 \item \textbf{RQ1: How does the performance of toxicity models change when psycholinguistic cues are added as features?}
 We found that using psycholinguistic features, we can achieve a slight improvement (by $\sim2\%$ in $F1_1$) in performance from baseline in identifying toxic instances across all classifiers.
 \item \textbf{RQ2: How does the performance of toxicity models change when moral values are added as features?}
 Adding morality on top of psycholinguistic features resulted in a significant jump in the performance of all classifiers ($\sim 2-7\%$ in $F1_1$).
 \item \textbf{RQ3: What types of SE texts are difficult to automatically detect as toxic using our techniques?}
 We conduct a qualitative error analysis to answer this question. Our observations provide several insights and potential areas of improvement to support future work in toxicity detection, e.g., using domain-specific dictionaries, understanding the context of the discussion, etc. 
\end{itemize}
\section{Background and Related Work}
\subsection{Moral Foundations Theory}
Moral Foundations Theory classifies human behavior into five core principles~\cite{graham2013moral}.  Each principle represents a pair of opposing values.
\textbf{Care/harm:} Rooted in our aversion to suffering—both for ourselves and others—this principle stems from our evolution as mammals, fostering virtues like kindness, compassion, and gentleness, while condemning cruelty and aggression.
\textbf{Fairness/cheating:} This principle emphasizes justice and rights, associated with the evolutionary concept of reciprocal altruism.
\textbf{Loyalty/betrayal:} Grounded in our tribal past, this principle encourages patriotism, heroism, trust, and self-sacrifice for the group, holding the ideal of “One for all, and all for one” as virtuous and viewing betrayal of social bonds as immoral.
\textbf{Authority/subversion:} Informed by hierarchical social structures in primate history, this principle upholds virtues like leadership, deference, and respect for authority and traditions, while perceiving challenges to authority as immoral.
\textbf{Sanctity/degradation:} Arising from the psychology of disgust and contamination, this principle views the body as a “temple” that can be defiled by immoral acts, promoting an elevated, noble approach to life.


\subsection{Toxicity in Online Communities}
Social media and online gaming communities are rife with online toxicity. Several factors contribute to online toxicity, including user anonymity, context collapse, and online disinhibition effect~\cite{Suler_2004}.
Different lexicons and annotated datasets have been used to study toxicity and abusive language on social media platforms.
State-of-the-art approaches such as classic machine learning models using features such as TF-IDF, part-of-speech tags, and sentiment to detect hateful content~\cite{davidson2017automated, founta2018large}, deep learning \cite{agrawal2018deep}, and transformer-based models \cite{isaksen2020using} have been used in toxicity detection. 
In addition, using LLMs for toxicity has shown promises but it has limitations in generalizability and understanding nuanced forms of toxicity~\cite{lee-etal-2024-improving, wang2022toxicitydetectiongenerativepromptbased}. Koh et al.~\cite{koh2024llmsrecognizetoxicitystructured} found that LLMs are unsuitable for blind toxicity evaluations within unverified factors.
Recent research showed that the majority of works and models used for toxicity detection ``encode'' biases against marginalized groups \cite{zhou2021challenges}. For instance, Sap et al.\cite{sap-etal-2022-annotators} found strong associations between toxicity rating and the identities and beliefs of human coders. Consequently, these biases are embedded in off-the-shelf models such as Perspective API, which are vastly used in this domain. 

\subsection{Toxicity in Open Source}
Similar to other online forums, OSS communication channels are not free of toxic content. 
Toxic behaviors have been responsible for causing stress and burnout~\cite{Raman2020}, reduced developer motivation and productivity~\cite{Bosu2013ImpactOP}, leading to team attrition~\cite{6614728}.
Systems to manage toxic comments have been used for several popular OSS platforms, e.g., GitHub.
Miller et al.~\cite{Miller2022} studied different aspects of toxicity on GitHub, including how developers react to the current moderation mechanism, noting that the current system does not always resolve the problem and that a significant amount of burden continues to be placed on the project maintainers. Therefore automated interventions to detect and flag toxic conversations in OSS are necessary.
Despite the presence of several state-of-the-art techniques for toxicity detection in blogs and tweets, applying these tools directly to the SE-related text is not effective due to several reasons such as longer texts containing references to code and other SE-specific factors~\cite{Sarker2020ABS, 9793879}. 
Towards domain-specific detection of toxic content, Raman et al. \cite{Raman2020} proposed a machine-learning-based technique to detect toxic issue comments on GitHub. Their model performed best when using only two features, the Stanford Politeness score and Google Perspective API. 
This model was found to be not generalizable across other types of developer communications, such as code reviews and chats~\cite{Sarker2020ABS}. 
Another study detected offensive language on Stack Overflow, GitHub, and chats by using the Perspective API and regular expressions~\cite{Cheriyan21} as features. LLMs also have shown promise in toxicity detection in OSS; however, they struggle with identifying passive-aggressiveness and context-dependent toxic language~\cite{10.1145/3639476.3639777}.


Current approaches in toxicity detection are not generalizable in terms that they miss classifying toxic/biased words that are salient and cultural or domain-specific (e.g., in SE context) and do not show up in off-the-shelf lexicons and datasets. Preliminary results from analyzing toxicity in OSS through the lens of moral values have shown promise in enhancing the understanding of toxic behaviors~\cite{10.1145/3611643.3613077}, and our work builds upon the previous works in this domain and investigates the change in the model's effectiveness when using additional features based on psycholinguistic scores and moral values.

\section{Methodology} We developed a suite of machine learning-based techniques for automatically identifying toxic SE communications on GitHub. Our approach takes as input a text segment, either an issue comment or a code review comment, and classifies them as toxic or non-toxic.

\begin{table}
\centering
\scriptsize
\caption{Overview of our experiment datasets}
\vspace{-0.2cm}
\label{tab:datasets}
\begin{tabular} {|c|c|c|c|c|}
\hline
   \textbf{Dataset} & \textbf{Source} & \textbf{\#Toxic} & \textbf{\#Non-toxic} & \textbf{Total}
  \\ \hline\hline

   Issue Comments & Github  & 101 & 303 & 404\\ \hline
   Code Review & Gerrit & 3,757 & 11,271 & 15028 \\ 
  \hline

  \hline
\end{tabular}
\vspace{-0.5cm}
\end{table}

\subsection{Datasets} \label{datasets}
We leverage two publicly available labeled datasets of toxic communications in the SE domain: (1) a dataset of 1,597 GitHub \textit{issue comments} (1,496 non-toxic, 101 toxic)~\cite{Raman2020}, and (2) a dataset of 19,571 Gerrit \textit{code review} comments (15,819 non-toxic, 3,757 toxic). The \textit{code review} dataset is an extension of \cite{Sarker2020ABS}. Both datasets are available in a public GitHub repository~\cite{gh_toxic_detector}.

The existing datasets are imbalanced, with less than 6\% and 20\% toxic instances in the \textit{issue comments} and \textit{code review} datasets, respectively. For our experiments, we select a representative subset of the original data with a ratio of 1:3 toxic to non-toxic instances. More specifically, we use undersampling techniques to reduce the size of the non-toxic class but retain all the toxic instances in our datasets. Table \ref{tab:datasets} shows the details of our datasets.

\begin{table}
\centering
\scriptsize
\caption{Features to identify toxicity}
\vspace{-0.2cm}
\label{tab:features}
\begin{tabular} {|c|c|c|}
\hline
    \textbf{Feature Set} & \textbf{Feature}  & \textbf{Value} 
    \\ \hline\hline

     Baseline &  Politeness  & 0 to +1 \\ 
        &  Google's Perspective API  &  0 to +1\\ \hline
     Psycholinguistic & LIWC (Analytic) &  1 to 99  \\ 
        & LIWC (Clout) &  1 to 99 \\ 
        & LIWC (Authentic) &   1 to 99\\ 
        & LIWC (Tone) &   1 to 99\\ 
        & LIWC (Swear) &   0 to 100\\ 
        & Sentiment &  -1 to +1 \\ \hline
      Morality & Care (Virtue and Vice) &  -1 to +1 \\ 
        & Fairness (Virtue and Vice) &  -1 to +1 \\ 
        & Ingroup (Virtue and Vice) &   -1 to +1\\ 
        & Authority (Virtue and Vice) &  -1 to +1 \\ 
        & Purity (Virtue and Vice) &  -1 to +1 \\ 
    \hline

    \hline
\end{tabular}
\vspace{-0.5cm}
\end{table}


\subsection{ML Classification} 
We investigated several supervised machine learning-based approaches to automatically identify toxic texts. We describe the textual features followed by the suite of machine learning algorithms investigated for this classification task.

\subsubsection{Features} 
We present three sets of features: Baseline (2 features), Psycholinguistic (6 features), and Morality (10 features). Table \ref{tab:features} lists the features in each set with their value range; descriptions of why and how we extract each feature follow.

\noindent
\textbf{Baseline Features:} 
The state-of-the-art domain-specific toxicity detector for SE \cite{Raman2020}, leverages Google's Perspective API~\cite{GooglePerspective} and Stanford's Politeness Detector~\cite{StanfordPoliteness}. Of the additional feature combinations \cite{Raman2020} experimented with (e.g., length of the text, subjectivity score, no. of anger words from LIWC), their model performed best when using only Politeness and toxicity score from the Perspective API. Hence, we use these two features in our baseline feature set. 


\noindent
\textbf{Psycholinguistic Features:}
To understand how toxicity is represented in text, we use a subset of features from the Linguistic Inquiry and Word Count (LIWC) \cite{boyd2022development}. Miller et al. found that the most prevalent forms of toxicity in OSS are entitled, trolling, arrogant, and unprofessional comments from project users, and insults arising from technical disagreements~\cite{Miller2022}. We use the following LIWC features to identify the presence or the lack of these traits: (a) \textit{Clout} for entitlement, (b) \textit{Authentic} for trolling, (c) \textit{Tone} for arrogance, (d) \textit{Analytic} for unprofessionalism, and (e) \textit{Swear words} for insults. These dimensions capture the linguistic and psychological cues relevant to OSS toxicity.

Sentiment is used for understanding people's emotions and affective states and is highly related to contentiousness \cite{pennacchiotti2010detecting}. 
In this work, we leverage Valence Aware Dictionary and sEntiment Reasoner (VADER) \cite{VADER} to get the sentiment scores of texts in our dataset. VADER performs well on short, informal content like OSS communications and is widely used in different domains, specifically SE~\cite{8453067}.
We used the compound score in VADER to get a single unidimensional measure of sentiment. The compound score sums the valence scores of each word in the lexicon and returns a ``normalized, weighted composite score'' for a given sentence.


\noindent
\textbf{Morality Features:}
As additional feature sets, we leverage Moral Foundations Theory~\cite{graham2013moral,graham2009liberals} to investigate the relationship between morality and toxicity. Our feature design is based on the premise that people's emotions, ideology, and culture can be reflected in their use of language~\cite{triandis1989self}. Hatred and tension (online or offline) may be the result of differences in values (moral or personal). Therefore, finding representations of such information in user-generated texts can help in better understanding toxicity in online interactions. 
The Moral Foundations Dictionary (MFD) enables the measurement of MFT based on text data by associating 324 words with virtues and vices from the MFT ~\cite{graham2009liberals,graham2013moral}.
To extract moral values, we used MFDE, an enhanced version of MFD~\cite{rezapour2019}. Compared to the original MFD, the enhanced lexicon consists of about 4,636 terms that were syntactically disambiguated and manually pruned and verified.
To find morality in texts, we apply Distributed Dictionary Representations (DDR) ~\cite{garten2018dictionaries}. DDR first computes the average of each dictionary and then computes the ``loading'' of a dictionary on a particular piece of text. We used MFDE as the seed words and created ten separate lists of words representing each moral value (vices and virtues of five moral categories). We used word2vec~\cite{mikolov2013efficient} to create vector representations of the words.

\subsubsection{Classifiers} We trained multiple supervised machine learning-based classifiers in our study using Python scikit-learn package~\cite{pedregosa2011scikit}. We explored other classifiers (e.g., Random Forests); however, we do not discuss them here, since they yielded significantly inferior results. Here, we provide an overview and explanation of our classifier choices.

\noindent
\textbf{Support Vector Machines (SVM)} is a non-probabilistic classifier that maps input data into a feature space that maximizes the gap between the classification categories; i.e., \textit{toxic}, and \textit{non-toxic} in our study. SVM has been observed to achieve high accuracy in predicting toxic content in SE~\cite{Raman2020, Cheriyan21}.

\noindent
\textbf{Logistic Regression (LR)} is a discriminative classification model that predicts the class by calculating the probability for each class and choosing the class with the highest probability. In our case, the class probability is the likelihood of a text being toxic. LR has been widely used for binary text classification in SE~\cite{Chatterjee2020Journal}.

\noindent
\textbf{Gradient Boosting (GB)} is an ensemble-based classification framework where a sequence of decision trees is constructed, and each tree minimizes the residual error of the preceding sequence of trees. 
Ensemble classifiers have been used in predicting offensive language in SE texts~\cite{Cheriyan21}.
\section{Evaluation Study}

\subsection{Evaluation Metrics} We use measures that are widely used for evaluation in information retrieval: precision, recall, F-measure, and ROC. To measure the fraction of automatically identified texts that are indeed toxic, we use precision, the ratio of true positives (TP) over the sum of true and false positives (FP). To see how often our approaches miss toxic data instances, we use recall, the ratio of true positives over the sum of true positives and false negatives (FN). F-measure combines these measures by harmonic mean. To measure robustness, we use ROC (Receiver Operating Characteristics), which represents the degree of separability between prediction classes. We compute precision, recall, and f-measure for non-toxic (class 0) and toxic (class 1) classes. 
Lastly, because our data is not completely balanced, we compute MCC (Matthews correlation coefficient), which is a correlation coefficient between observed and predicted binary classifications that is well suited for unbalanced data. 

\subsection{Procedures} We configured classifiers and ran them as follows.

\subsubsection{Hyperparamter Tuning} We investigated several hyperparameters to adjust each classifier (using scikit-learn), and the following configurations produced the best classifications. For the rest of the parameters, we used the default values offered by scikit-learn.

\begin{itemize}[leftmargin=*]
\item Support-Vector Machine (SVM): We used the \textit{LinearSVC} class with the following parameters: C=10, max\_iter=10000.

\item Logistic Regression (LR): We used the \textit{LogisticRegression} class with following parameters: C=1, max\_iter=4000, multi\_class='multinomial'.

\item  Gradient-Boosting (GB): We used the \textit{GradientBoostingClassifier} class with the following parameters: learning\_rate=1.0, n\_estimators=1000, random\_state = 0, max\_depth=10, max\_features='sqrt', min\_samples\_leaf=2. 
\end{itemize}


\begin{table*}[t]
\centering
\caption{Toxicity Detection Results (Issue Comments) (5-Fold Cross Validation)}
\vspace{-0.2cm}
\resizebox{\textwidth}{!}{%
\begin{tabular}{|l|l|c|c|c|c|c|c|c|c|c|}
\hline
 &  & \textbf{$P_0$} & \textbf{$R_0$} & \textbf{$F1_0$} & \textbf{$ROC_0$} & \textbf{$P_1$} & \textbf{$R_1$} & \textbf{$F1_1$} & \textbf{$ROC_1$} & \textbf{$MCC$} \\ \hline\hline
{\textbf{Baseline}} & \textit{LinearSVC} & 84.13 & 97.00 & 90.09 & 71.29 & 84.92 & 45.57 & 58.87 & 71.29 & 54.05 \\ 
 & \textit{GradientBoosting} & 86.24 & 87.67 & 86.94 & 73.02 & 61.48 & 58.38 & 59.83 & 73.02 & 46.87 \\ 
 & \textit{LogisticRegression} & 83.34 & 98.00 & 90.05 & 69.83 & 88.89 & 41.67 & 56.08 & 69.83 & 53.26 \\ 
 \hline\hline
{\textbf{Baseline+Psycholinguistic}} & \textit{LinearSVC} & 84.52 & 96.33 & 90.03 & 71.95 & 82.52 & 47.57 & 60.03 & 71.95 & \textbf{54.14} \\ 
 & \textit{GradientBoosting} & 86.27 & 92.00 & 89.01 & 74.24 & 71.97 & 56.48 & \textbf{62.75} & \textbf{74.24} & 52.98 \\ 
 & \textit{LogisticRegression} & 83.72 & 97.67 & 90.15 & 70.62 & 86.91 & 43.57 & 57.86 & 70.62 & 53.89 \\ 
\hline\hline
{\textbf{Baseline+Psycholinguistic+Morality}} & \textit{LinearSVC} & 87.50 & 90.33 & 88.79 & 75.90 & 70.59 & 61.48 & 64.74 & 75.90 & \textbf{54.64} \\
 & \textit{GradientBoosting} & 87.65 & 89.67 & 88.56 & 76.02 & 68.69 & 62.38 & \textbf{64.83} & \textbf{76.02} & 54.06 \\ 
 & \textit{LogisticRegression} & 84.85 & 95.00 & 89.62 & 72.26 & 77.36 & 49.52 & 60.13 & 72.26 & 52.55 \\ 
\hline
\end{tabular}
}
\label{tab:issue-comments-classifier}
\vspace{-0.4cm}
\end{table*}

\begin{table*}[t]
\centering
\caption{Toxicity Detection Results (Code Review) (5-Fold Cross Validation)}
\vspace{-0.2cm}
\resizebox{\textwidth}{!}{%
\begin{tabular}{|l|l|c|c|c|c|c|c|c|c|c|}
\hline
 &  & \textbf{$P_0$} & \textbf{$R_0$} & \textbf{$F1_0$} & \textbf{$ROC_0$} & \textbf{$P_1$} & \textbf{$R_1$} & \textbf{$F1_1$} & \textbf{$ROC_1$} & \textbf{$MCC$} \\ \hline\hline
{\textbf{Baseline}} & \textit{LinearSVC} & 84.11 & 89.93 & 86.93 & 69.42 & 61.82 & 48.91 & 54.60 & 69.42 & 42.24 \\ 
 & \textit{Gradient Boosting} & 81.78 & 75.01 & 77.95 & 62.75 & 42.49 & 50.50 & 45.48 & 62.75 & 24.85 \\ 
 & \textit{Logistic Regression} & 84.15 & 89.90 & 86.93 & 69.49 & 61.81 & 49.08 & 54.71 & 69.49 & 42.32 \\ 
\hline\hline
{\textbf{Baseline+Psycholinguistic}} & \textit{LinearSVC} & 83.62 & 95.21 & 89.04 & 69.55 & 75.24 & 43.88 & 55.42 & 69.55 & \textbf{47.96} \\ 
 & \textit{Gradient Boosting} & 87.10 & 79.67 & 82.91 & 72.76 & 57.04 & 65.85 & \textbf{60.09} & \textbf{72.76} & 44.73 \\ 
 & \textit{Logistic Regression} & 84.39 & 92.73 & 88.36 & 70.56 & 68.91 & 48.40 & 56.85 & 70.56 & 46.82 \\ 
\hline\hline
{\textbf{Baseline+Psycholinguistic+Morality}} & \textit{LinearSVC} & 84.87 & 94.58 & 89.46 & 71.92 & 75.41 & 49.25 & 59.46 & 71.92 & 51.37 \\  
 & \textit{Gradient Boosting} & 88.60 & 88.25 & 88.07 & 77.35 & 71.09 & 66.45 & \textbf{67.50} & \textbf{77.35} & \textbf{57.03} \\ 
 & \textit{Logistic Regression} & 85.82 & 93.00 & 89.25 & 73.35 & 72.01 & 53.70 & 61.43 & 73.35 & 51.94 \\ 
\hline
\end{tabular}
}
\label{tab:code review-classifier}
\vspace{-0.5cm}
\end{table*}

\subsubsection{Evaluation Process (RQ1, RQ2)} For RQ1 and RQ2, results from the classifiers were obtained using stratified 5-fold cross validation i.e., the dataset was partitioned into five equal-sized sub-samples with stratification, ensuring that the original distribution of toxic and non-toxic instances is retained in each sub-sample.

\subsubsection{Evaluation Process (RQ3)}
For RQ3, we developed two separate sets of test data from issue comments and code reviews to evaluate the effectiveness of our models, and investigate the properties of SE texts that were found to be challenging to automatically categorize.
For the code review test set, we randomly selected 100 toxic and 100 non-toxic reviews from the original dataset, before creating the training described in the Methodology section. 
For the second dataset, since the issue comments data is small in size and only consists of 101 toxic instances, we additionally utilized the held out labeled issue test set created by \cite{Raman2020}, which consists of 194 issue threads labeled as toxic or non-toxic. Since the issue threads tend to be longer compared to our training data, we used the length of texts to filter out threads longer than 1,700 characters to have data similar to our training set. Our final issue comments test set consists of 58 issue comments (25 toxic, 33 non-toxic).

\subsection{Results and Discussion}
Tables \ref{tab:issue-comments-classifier} and \ref{tab:code review-classifier} present the precision, recall, F-measure, ROC, and MCC, for each classification labels (0 for non-toxic, and 1 for toxic). To configure the feature sets, we first combine baseline features with psycholinguistic features (baseline + psycholinguistic), and then combine baseline features with both psycholinguistic and morality features (baseline + psycholinguistic + morality). We compare the baseline with both of these configurations. 

\paragraph{\textbf{RQ1.}  How does the performance of toxicity models change when psycholinguistic cues are added as features?}

When using \linguistic{} features, we observe an improvement in performance from baseline 
in identifying toxic instances across all classifiers. 

In the issue comments dataset (Table \ref{tab:issue-comments-classifier}), the $F1_1$ improves from 58.87\% to 60.03\% for \textit{LinearSVC}, from 59.83\% to 62.75\% for \textit{GradientBoosting}, and from 56.08\% to 57.86\% for \textit{LogisticRegression}. When considering $ROC_1$ and $ROC_2$, we observe slight improvement ($\sim1\%$) from the baseline, across all classifiers. Overall (except MCC), \textit{GradientBoosting} performs better compared to the rest of the classifiers. However, when considering $MCC$, \textit{LinearSVC} provides the best performance (54.14\%).  

In the code review dataset (Table \ref{tab:code review-classifier}), the $F1_1$ improves from 54.60\% to 55.42\% for \textit{LinearSVC}, from 45.48\% to 60.09\% for \textit{GradientBoosting}, and from 54.71\% to 56.85\% for \textit{LogisticRegression}. Overall (except MCC), \textit{GradientBoosting} performs better than the rest of the classifiers; we observe a significant improvement from the baseline with $\sim$15\% (45.48\% to 60.09\%) in $F1_1$, $\sim$10\% (62.75\% to 72.76\%) in $ROC_1$. 
However, for $MCC$ that adjusts for class imbalance, \textit{LinearSVC} provides better performance (47.96\%). 
\textit{LinearSVC} also provides a high precision ($P_1$ = 75.24\%), but a lower recall ($R_1$ = 43.88\%), indicating that it is more restrictive in labeling a code review comment as toxic.

We further looked into the values of psycholinguistic features in both toxic and non-toxic comments in our datasets. In both datasets, the average value of ``Swear words'' is significantly higher in toxic comments, 2.72 vs. 0.01 in toxic and non-toxic comments in issue comments, and 6.3 vs. 0.11 in the code review dataset. We also observe that, the average value of ``Analytic" is lower in toxic comments, 30.96 vs 52.3 in toxic and non-toxic comments in issue comments, and 40.99 vs 44.8 in the code review dataset. Since, analytic scores represent the degree of formal, logical, and hierarchical thinking~\cite{Pennebaker_2015}, a higher value shows that non-toxic comments were well-thought-out and thus exhibited more professionalism. Additionally, we found that the average value of ``Authentic" scores is higher in toxic comments. Example of texts that have low authenticity scores include texts where a person is being socially cautious~\cite{Newman2003} and thus not involved in toxic behaviors such as trolling. These results also confirm \cite{Miller2022}'s empirical observation of the existence of trolling, insults, and unprofessional comments as prevalent forms of OSS toxicity. 
We do not observe any significant patterns with the rest of the psycholinguistic features across both the datasets, which indicates the challenges of generalizing toxicity detection techniques for different types of data.

\paragraph{\textbf{RQ2.} How does the performance of toxicity models change when moral values are added as features?}
Adding morality on top of \linguistic{} resulted in a significant jump in the performance of all classifiers. 

In the issue comments dataset (Table \ref{tab:issue-comments-classifier}), while \textit{LinearSVC} benefits the most from the addition of moral values, with around 4\% increase in the $F1_1$, \textit{GradientBoosting} achieves the highest performance of all models ($F1_1 = 64.83$\%). Furthermore, using this model with \morality{} results in the highest $ROC_1$ as well as $ROC_0$. When considering $MCC$, we observe a slightly better result with \textit{LinearSVC}. 
In the code review dataset (Table \ref{tab:code review-classifier}), the \textit{GradientBoosting}'s performance jumps from 60.09\% with \linguistic{} to 67.5\%. $MCC$, $ROC_1$, and $ROC_0$ also significantly improved using this model. 

We further looked into the morality features in both toxic and non-toxic comments and found that the average of all moral values are higher in toxic comments compared to the non-toxic ones in both datasets used in this paper. In both datasets, the average value of ``degradation'' (purity-vice) is significantly higher, $0.41\pm 0.05$ vs. $0.35\pm 0.08$ in toxic and non-toxic comments in issue comments, and $0.380\pm 0.07$ vs. $0.34\pm 0.08$ ones in the code review dataset. 

Based on MFT, purity/degradation foundation is influenced by the ``psychology of disgust and contamination''. The result of our analysis is in line with previous work that found hate is conceptually closer to disgust and contempt, compared to anger and dislike \cite{martinez2022hate}. 

\paragraph{\textbf{RQ3.} What types of SE texts are difficult to automatically detect as toxic using our techniques?}
To answer RQ3, we performed classification error analysis. Specifically, we investigated the following questions:

\vspace{0.2cm}

\noindent
\textbf{RQ3.1.} \textit{What instances are misclassified using all feature sets?}

\noindent
We qualitatively analyzed the False Positives (FP) and False Negatives (FN) using all features, and GB as the classifier. We chose GB since it achieves the best classification performance overall. We found that 30 instances (4 issue comments, 26 code reviews) were marked as FP, and 45 instances (7 issue comments, 38 code reviews) were marked as FN, out of a total of 258 instances in our evaluation dataset.

The analysis procedure consisted of the following steps: (1) first we collected the data instances that were marked FP, and instances that were marked FN, using all features. (2) Following an open coding procedure \cite{Runeson:2012}, the authors of this paper independently studied the instances from step 1. We manually analyzed the conversations to identify the characteristics of conversations in each category (FP and FN), and recorded comments and reflections from the manual analysis in the form of short phrases, e.g., ``contains entire sentence(s) written in uppercase''. These insights helped us investigate additional characteristics that our features failed to capture. (3) The common observed characteristics in each category (FP and FN) were grouped. The analysis was performed in an iterative approach composed of multiple sessions, which helped in generalizing the hypotheses and revising the characteristics. 

We manually analyzed the 30 instances marked FP when using all features, and observed that most of the instances contained domain-specific words that were incorrectly identified as toxic. For instance, \textit{``If we add this board, will we have to start killing off the rest?..."}, where `killing' refers to terminating an application. These errors could be handled using a domain specific dictionary. Other errors included misidentifying self-deprecating words as toxic. For instance, \textit{``Oops, that was dumb. I actually had caught this while you were reviewing and fixed it."}. This is a limitation of using lexicon-based features, as words can have different meanings in different contexts. 

We manually analyzed the 45 instances marked FN when using all features, and observed that toxicity in several of the instances could be understood only in the context of the discussion and with prior project-specific knowledge. For instance, \textit{``extra stupid-people safe: do we want `pwd -P' here, in case someone runs this in a symlinked directory? :)"}. The other toxic comments were insults without containing any negative terms such as swear words. For instance, in the comment   \textit{``...The questions are placed looking for fixes not closed stamps. Did they give you that stamp in Kindergarten?..."}, the words ``stamps" and ``Kindergarten" refer to a false sense of achievement.  
We also notice that some comments included entire sentences written in caps (or uppercase), which insinuates shouting or harsh tone, e.g., \textit{``...@user already said why, stop asking.  BUT CANT YOU JUST CHANGE BESTPLANET IN THE CODE AND SET IT TO PLANET 22..."}. These instances are misclassified by our models, since we are not leveraging the format of the text as a feature. 

\vspace{0.2cm}

\noindent
\textbf{RQ3.2.} \textit{What instances are misclassified by other feature sets but not morality?}

\noindent
Overall, we observed the best performance when all features are used; specifically adding morality features significantly improves the performance (5\% with SVM for issue comments, 7\% with GB on code reviews). In this question, we investigate and gather insights on how adding morality features improves the classification of toxicity in OSS. 

We selected the instances which were misclassified by GB classifier using only baseline and psycholinguistic features, but correctly classified when all feature sets are used (including morality). Using this selection procedure, we collected a set of 33 FPs (3 issue comments, 30 code reviews), and 22 FNs (2 issue comments, 20 code reviews). The analysis procedure was the same as RQ3.1.

We qualitatively analyzed the 33 FPs, and observed that the majority of comments contain domain-specific words. For instance, \textit{``if you call  response.body()
.string() twice like  this response.body().string() response.body().string() you will get this unsolvable error''} were misclassified using \linguistic{} features, but adding morality on top of that reduced model confusion. Another instance is \textit{``Bad merge? You're shadowing oslo.config.cfg''} which was mistakenly classified as toxic using the baseline and \linguistic{} features. However, using moral words in addition to other features complemented each other, as also shown in other studies \cite{garten2018dictionaries}, which resulted in decreasing classification error in our models.

We qualitatively analyzed the 22 FNs, and observed that the majority of comments misclassified by other models include negative words such as ``damn'', ``stupid'', and ``ugly'' which are not included as swear words in LIWC. One of the shortcomings of LIWC is that the analysis is on the word level, which results in missing semantically related words in many cases. Using morality as a feature and the DDR method helped with capturing semantic similarity between words and concepts and going beyond word count. For instance, a comment \textit{``this is pretty ugly. create it in one assignment''} is classified as non-toxic by the baseline and \linguistic{} features, but using moral words improved the performance of the model and resulted in a better accuracy. 

\vspace{0.2cm}
\noindent
\textbf{Summary of Key Findings and Takeway. } 
We found that psycholinguistic markers, such as ``Swear words" and ``Analytic" scores, proved valuable in distinguishing toxic from non-toxic content, while moral values (particularly ``Degradation") captured nuanced forms of toxicity often linked to moral judgments. These features can be leveraged to build more accurate toxicity detectors for SE domains.

However, several challenges persist. SE-specific terms and indirect toxicity, including sarcasm, entitlement, and passive-aggressiveness, are often misclassified without domain-aware dictionaries and contextual understanding. Additionally, the complexity of OSS communication requires models that can account for conversation dynamics and project-specific knowledge. Future research should focus on refining these features and developing models capable of handling covert toxicity in diverse SE contexts, potentially by incorporating contextual markers and adaptive, SE domain-specific lexicons.




\section{Conclusion and Future Work}
This paper investigates the usefulness of psycholinguistic cues and moral values with toxicity in Open Source Software (OSS). Due to the use of domain-specific words and jargon in OSS, the majority of toxicity detection tools do not correctly identify toxic language in this domain. Therefore, there is a need to augment the current methods with features and information that are insightful for such task. 
A recent study~\cite{Miller2022} in OSS found features such as trolling, arrogant, and unprofessional comments mostly prevalent to toxicity in OSS. 
In this work, we leverage Linguistic Inquiry and Word Count (LIWC) to map these concepts to a subset of linguistic features in comments labeled as toxic and non-toxic. 
In addition, previous studies showed that people's instant emotions and ideologies can be reflected in their use of language. Moral Foundations Theory captures the reactions of people. We use an extended version of Moral Foundations Dictionary \cite{rezapour2019} to operationalize morality in text. 
Using these two sets of additional feature sets on top of a benchmark baseline model, we show that using moral values on top of psycholinguistic features improves the performance of our toxicity classifiers. 

This study opens avenues for enhancing toxicity detection in OSS. Our scripts are available in our replication package~\cite{anonymized}. Future work can focus on building larger, more balanced datasets to improve the generalizability of toxicity models across diverse OSS contexts. Additionally, moving beyond binary classification toward multi-dimensional toxicity categorization (e.g., personal harassment, implicit bias) will capture the complexity of toxic interactions, particularly those affecting underrepresented groups~\cite{9261329}. Additionally, developing models that understand the evolution of conversations—recognizing how tone, sarcasm, and passive-aggressiveness unfold over exchanges—could improve the detection of covert toxicity. 



\bibliographystyle{IEEEtran}
\bibliography{aaai22, oss, preetha}

\end{document}